\documentclass[preprint]{elsarticle}
\usepackage{amsmath}
\usepackage{amssymb}
\usepackage{graphicx}
\usepackage{float}
\newcommand{\wt}[1]{\widetilde{#1}}
\newcommand{\eq}[1]{#1^{(eq)}}

\newcommand{\req}[1]{Eq.\,({\ref{#1}})}

\newcommand{\reff}[1]{Fig.\,({\ref{#1}})}

\begin{document}
%\preprint{Draft.13}
\title{Electron-positron plasma in BBN:\\ damped-dynamic screening}
\author{Christopher Grayson$^a$, Cheng Tao Yang$^a$, \\Martin Formanek$^b$, Johann Rafelski$^a$}
\affiliation{Department of Physics, The University of Arizona, Tucson, Arizona 85721, USA}
\affiliation{ELI Beamlines Facility, The Extreme Light Infrastructure ERIC, 252 41 Dolni Brezany, Czech Republic}
\date{August 17, 2023}

%%%%%%%%%%%%%%%%%%%%%%%%%%%%%%%%%%%%%%%%%%%%%%%%%%%%% 
\begin{abstract}
We characterize in detail the dense $e^- e^+ \gamma$ plasma present during the Big-Bang Nucleosynthesis (BBN) and explore how it is perturbed electromagnetically by \lq\lq impurities, {\it i.e.\/}, spatially dispersed protons and light nuclei undergoing thermal motion. The internuclear electromagnetic screened potential is obtained (analytically) using the linear response approach, allowing for the dynamic motion of the electromagnetic field sources and the damping effects due to plasma component scattering. We discuss the limits of the linear response method and suggest additional work needed to improve BBN reaction rates in the primordial Universe. Our theoretical methods describing the potential between charged dust particles align with previous studies on planetary and space dusty plasma and could have a significant impact on the interpretation of standard cosmological model results.
 
\end{abstract}
\maketitle

%%%%%%%%%%%%%%%%%%%%%%%%%%%%%%%%%%%%%%%%%%%%%%%%%%%%%%%%%%%%%%%%%%%
\section{Introduction}\label{sec:intro}

Electron-positron $e^-e^+$-pair-plasma has a significant presence in the early Universe during Big-Bang nucleosynthesis (BBN)~\cite{ Wang:2010px, Hwang:2021kno, Rafelski:2023emw}. Charge screening is one of the important collective plasma effects modifying the internuclear potential $\phi(r)$. The electron cloud surrounding the charge of an ion screens other nuclear charges far from their own radius and reduces the Coulomb barrier. In nuclear reactions, reducing the Coulomb barrier increases the penetration probability, enhances thermonuclear reaction rates, and thus modifies light element abundances in the early universe. 

The screening effect in nuclear reactions has a long and distinguished history. Plasma screening effects were first considered in 1954 by Salpeter~\cite{Salpeter:1954nc}, who proposed evaluating the enhancement of nuclear reactions by employing the static Debye-H\"{u}ckel potential~\cite{Debye:1923,Salpeter:1969apj,Famiano:2016hhs}. These applications focus on collision-less electron-positron plasma only. Later, this approach was generalized for nuclei moving in the plasma~\cite{Hwang:2021kno,Carraro:1988apj, Gruzinov:1997as, Opher:1999jh, Yao:2016cjs} {\it i.e.\/}, `dynamic' screening. In this work, we address the ultra dense nature of the quantum electrodynamics (QED) $e^-e^+\gamma$ plasma in which the BBN reaction network occurs. Only recently has the ultra dense nature of the BBN plasma been taken into account by the inclusion of collisions in numerical simulations \cite{Sasankan:2019oee,Kedia:2020xdc}. The dense nature of the the BBN plasma is important since the BBN reactions typically take place within the temperature interval $86\, \mathrm{keV}>\mathrm{T_{BBN}}>50\, \mathrm{keV}$~\cite{Pitrou:2018cgg} which is a non-negligible fraction of the electron mass energy equivalent, $m_ec^2=511$\,keV. The individual densities of electron-positron pairs are higher than the electron density in the solar core and surpass densities found in most stellar burning environments. This dense plasma environment indicates the need to understand the effect of damping due to scattering of plasma components $e^- e^+ \gamma$ during the BBN epoch.

To improve the understanding of the internuclear screened Coulomb potential $\phi$ governing BBN reaction rates, this work expands the prior studies~\cite{Hwang:2021kno, Carraro:1988apj,Opher:1999jh, Yao:2016cjs} carried out in the linear response approximation applicable in the regime where the modification of the Coulomb potential is a small effect. We extend the 'dynamic screening' effort (and in particular that of Hwang et al~\cite{Hwang:2021kno} which introduced dynamic screening during BBN) to study `damped-dynamic' screening as follows:
\begin{enumerate}
\item 
We detail the scattering damping effect of plasma components in the temperature range beginning near $T\simeq m_ec^2=511$\,keV extending down to a temperature near $T_\mathrm{split}\simeq 20.3$\,keV where practically all $e^-e^+$-pairs disappeared.
\item 
We obtain analytical results for damped-dynamic screening applicable to the BBN epoch temperature range which demonstrates a connection between the BBN plasma and dusty plasma theory. 
\item 
We recognize and describe the limits of applicability of the linear response method to BBN by considering the full equilibrium distribution necessary in the presence of strong fields. We explore a one-dimensional toy model at distances a fraction of an \AA ngstrom in the quantum tunneling regime where the potential $\phi$ is large compared to the thermal collision energy $|\phi|/3T> 1$,  which predicts enhanced screening at small distances.
\end{enumerate}

Such effort to improve the BBN model through screening is well justified for the following reasons:
\begin{enumerate}
\item Currently, the most accessible path to observe the early Universe before recombination is through light element abundances. 
\item
Despite the great initial success of the BBN model regarding the prediction of the abundances of D, $^3$ He, $^4$ He, there remain well-documented discrepancies. Currently, substantial efforts are being directed toward reconciling the theoretical BBN model with present-day observations. For a more detailed discussion, see Refs.\,\cite{Pitrou:2021vqr,Bertulani:2022qly}.
\item
Enhanced production of light elements during the BBN era could help resolve two problems: a) Elements Beryllium and Boron cannot be produced in stellar burning processes while their high abundance suggests additional mechanisms of production beyond secondary processes after stellar explosions. b) Understanding the age of stars obtained using the abundance of light elements (metallicity) may be improved by adaptation of the BBN network to account for nuclear reactions in ultra-dense QED plasma.
\item
We note that screening effects are more pronounced for nuclei with greater charge $Z$ leading to stronger modification in predicted abundances, just where the most significant current disagreements exist ({\it e.g.\/} the abundance of $^7$Li). Inclusion of screening effects in our current models of the BBN reaction network could explain this discrepancy without the need to invoke more exotic modifications of the Coulomb internuclear potential. An example of such a modification would be the variation of $\alpha$, the fine-structure constant~\cite{Meissner:2023voo}.
\item 
Recent James Webb space telescope-Hubble space telescope (JWST-HST) observations implying early onset of galactic structure formation and/or presence of luminous objects as early as 300 million years after BBN~\cite{Haro:2023JWST,Sabti:2023xwo,Ilie:2023DM} could be seen as evidence supporting nuclear dust matter clustering in dense QED plasma.
\end{enumerate}

In the standard BBN model, thermonuclear reaction rates are evaluated using nuclear reactions specific to the vacuum state. In a high-precision BBN model, the effect of $e^-e^+$-pair-plasma screening of nuclear charges must be allowed. We describe in Section~\ref{sec:density} the presence of up to several millions of electron-positron ($e^-e^+$) pairs per every charged nucleon. This situation prompts an effort to reconsider BBN in an ultra dense plasma environment allowing for a dense $e^-e^+\gamma$ plasma medium.

We account for collisions between plasma components using the current-conserving Bhatnagar, Gross, and Krook (BGK) collision term~\cite{Bhatnagar:1954zz}, which allows us to study damping in the dynamic plasma. For an in-depth discussion presented in contemporary covariant notation and preserving current conservation, see Formanek et al.~\cite{Formanek:2021blc}. Our approach is different from Monte-Carlo simulation of two particle collisions~\cite{Sasankan:2019oee,Kedia:2020xdc} as in our case particles participating in scattering also experience simplified static screening which in terms of collision description would require more than two particle collisions. In addition, the simplified BGK collision term we employ allows us to obtain an analytic `damped-dynamic' internuclear potential relevant to the BBN reaction network in the linear response approximation. 

Several well known processes, including M\o ller, Bhabha, and inverse Compton scattering, characterize the damping in plasma. These textbook results will be adapted to the plasma environment: When studied in vacuum, electrons and positrons interact with each other by exchanging a massless photon. However, when a photon propagates through a plasma of electrons and positrons, its properties are modified by interactions with the medium. When evaluating Moller and Bhabha scattering, we include the temperature-dependent mass of the photon obtained in plasma theory without damping. We find that the total relaxation rate during BBN is much larger than the screening mass, which is the inverse of the Debye screening length $m_D = 1/\lambda_D$. This suggests that electromagnetic perturbations of the plasma will be overdamped.

Since, as noted above, the computed damping strength is the dominant scale, it is also the main parameter determining the photon mass. However, introducing the damping strength in the photon mass would make the calculation of the damping strength self-referential. The complexity of such a self-consistent evaluation of damping is beyond the scope of this work. This underscores the need to develop a self-consistent approach where both damping and photon properties in plasma are determined in a mutually consistent manner.

The QED $e^+e^-\gamma$-plasma present during BBN is perturbed by a variety of comparatively very heavy charged protons $p$ and even heavier isotopes and light elements. In calculating the electromagnetic potential $\phi$, we naturally view BBN active participants as heavier, higher charged impurities {\it i.e.\/} \lq\lq charged dust\rq\rq in the QED-plasma. Since theories of such a system have been developed by others working on charged dust grains in planetary and space plasma~\cite{Montgomery:1970jpp, Stenflo:1973, Shukla:2002ppcf, Lampe:2000pop} we can validate our results and adapt previously developed theoretical tools necessary for describing heavy charged impurities within a plasma of lighter particles.

We provide the first step in merging these two fields by including damping to analyze the electromagnetic potential during BBN, similar to work done by Stenflo in 1973 for dusty plasma~\cite{Stenflo:1973}. We advance this work by applying it to the BBN epoch, allowing for the finite size of the "dust grains," and by finding the short-distance behavior of the potential in the weak field limit. Using a general model of simplified collisional damping in linear response, we calculate the screened electromagnetic potential of light nuclei undergoing thermal motion subject to damping relying on previous work ~\cite{Formanek:2021blc,Grayson:2022asf}.

Our manuscript is organized as follows: In Section~\ref{sec:density} we derive the electron-positron density and chemical potential to show that during the normal BBN temperature range $86\,\mathrm{keV}>\mathrm{T_{BBN}}>50\,\mathrm{keV}$~\cite{Pitrou:2018cgg} the Universe was filled with a dense electron-positron pair-plasma dotted with dispersed baryonic matter dust. 
In Section~\ref{sec:relax}, we calculate the relaxation rate $\kappa= 1/\tau$, where $\tau$ is the mean time between collisions in the plasma. This is done by finding the average of the most relevant reaction rates for $2\leftrightarrow 2$ scattering: M{\o}ller, Bhabha, and inverse Compton scattering. 

We proceed in Section~\ref{sec:kinetic_theory} to solve the Vlasov-Boltzmann Equation (VBE) in the weak field limit with damping. Our approach considers small plasma perturbations in linear response to find the polarization tensor present during the BBN era. In Section~\ref{sec:potential}, we find an analytic expression for the damped-dynamic potential applicable to BBN and study the limitation of the linear response method. We show that additional strong field phenomena beyond linear response alter the behavior of the screened potential $\phi$. We discuss our results and describe their interdisciplinary importance in Section~\ref{sec:Discussion}. 
 
%%%%%%%%%%%%%%%%%%%%%%%%%%%%%%%%%%%%%%%%%%%%%%%%%%%%%%%%%%%%%%%%%%%

\section{Electron chemical potential and number density}\label{sec:density}
In this section, we will derive the dependence of electron chemical potential, and hence $e^+e^-$ density, as a function of the photon background temperature $T$ by employing the following physical principles:
\begin{enumerate}
\item Charge neutrality of the Universe:
\begin{align}\label{neutrality}
n_{e^-}-n_{{e^+}}=n_p-n_{\overline{p}}\approx\,n_p,
\end{align}
where $n_\ell$ denotes the number density of particle type $\ell$.
\item Neutrinos decouple (freeze out) at a temperature $T_f\simeq 2$ MeV, after which they free stream through the Universe with an effective temperature~\cite{Birrell:2012gg}
\begin{align}
T_\nu(t)=T_f\,\frac{a(t_f)}{a(t)},
\end{align}
 where $a(t)$ is the--Friedmann--Lema\^{i}tre--Robertson--Walker (FLRW) Universe scale factor which is a function of cosmic time $t$, and $t_f$ represents the cosmic time when neutrino freezes out.
\item Total comoving entropy is conserved. At $T\leq T_f$, the dominant contributors to entropy are photons, $e^+e^-$, and neutrinos. In addition, after neutrino freeze out, neutrino comoving entropy is independently conserved~\cite{Birrell:2012gg}. This implies that the combined comoving entropy in $e^+e^-\gamma$ is also conserved for $T\leq T_f$.
\end{enumerate}

Motivated by the fact that comoving entropy in $\gamma$, $e^+e^-$ is conserved after neutrino freezeout, we rewrite the charge neutrality condition, Eq.~(\ref{neutrality}), in the form
\begin{align}\label{charge_neutral_cond2}
n_{e^-}-n_{{e^+}}=X_p\frac{n_B}{s_{\gamma,e^\pm}} s_{\gamma,e^\pm},\qquad X_p\equiv\frac{n_p}{n_B},
\end{align}
where $n_B$ is the number density of baryons, $s_{\gamma,e^\pm}$ is the combined entropy density in photons, electrons, and positrons. During the Universe expansion, the comoving entropy and baryon number are conserved quantities; hence the ratio $n_B/s_{\gamma,e^\pm}$ is conserved. We have
\begin{align}
\frac{n_B}{s_{\gamma,e^\pm,}}=\left(\frac{n_B}{s_{\gamma,e^\pm}}\right)_{t_0}\!\!\!\!=\left(\frac{n_B}{s_{\gamma}}\right)_{t_0}\!\!\!\!=\left(\frac{n_B}{n_\gamma}\right)_{t_0}\left(\frac{n_\gamma}{s_{\gamma}}\right)_{t_0},
\end{align}
where the subscript $t_0$ denotes the present day value, and the second equality is obtained by observing that the present day $e^+e^-$-entropy density is negligible compared to the photon entropy density. We can evaluate the ratio by giving the present day baryon-to-photon ratio: $B/N_\gamma =n_B/n_\gamma= 0.605\times10^{-9}$ from Cosmic Microwave Background (CMB)~\cite{ParticleDataGroup:2022pth} and the entropy per particle for a massless boson: $(s/n)_{\mathrm{boson}}\approx 3.602$.

The total entropy density of photons, electrons, and positrons can be written as
\begin{align}\label{entropy_per_baryon}
s_{\gamma,e^\pm}=\frac{2\pi^2}{45}g_\gamma\,T^3+\frac{\rho_{e^\pm}+P_{e^\pm}}{T}-\frac{\mu_e}{T}(n_{e^-}-n_{{e^+}}),
\end{align}
where $ \rho_{e^\pm}=\rho_{e^-}+\rho_{e^+}$ and $P_{e^\pm}=P_{e^-}+P_{{e^+}}$ are the total energy density and pressure of electrons and positron respectively.

By incorporating Eq.~(\ref{charge_neutral_cond2}) and Eq.~(\ref{entropy_per_baryon}), the charge neutrality condition can be expressed as
\begin{align}\label{charge_neutral_cond3}
&\left[1+X_p\left(\frac{n_B}{n_\gamma}\right)_{t_0}\left(\frac{n_\gamma}{s_{\gamma}}\right)_{t_0}\frac{\mu_e}{T}\right]\frac{n_{e^-}-n_{{e^+}}}{T^3}\notag\\
&\qquad\qquad\qquad=X_p\left(\frac{n_B}{n_\gamma}\right)_{t_0}\left(\frac{n_\gamma}{s_{\gamma}}\right)_{t_0} \left(\frac{2\pi^2}{45}g_\gamma+\frac{\rho_{e^\pm}+P_{e^\pm}}{T^4}\right).
\end{align}

Using Fermi distribution, the number density of electrons over positrons in the early Universe is given by
\begin{align}\label{ee_density}
n_{e^-}-n_{{e^+}}&=\frac{g_e}{2\pi^2}\left[\int_0^\infty\frac{p^2dp}{\exp{\left((E-\mu_e)\right)/T}+1}\right.\left.-\int_0^\infty\frac{p^2dp}{\exp{\left((E+\mu_e)/T\right)}+1}\right]\notag\\
&=\frac{g_e}{2\pi^2}\,{T^3}\,\tanh(b_e)M_e^3\int_{1}^\infty \!\!\!\!\frac{ \eta \sqrt{\eta^2-1} d\eta}{1+\cosh(M_e\eta)/\cosh(b_e)},
\end{align}
where we have introduced the dimensionless variables as follows: 
\begin{align}\label{Variables}
\eta=\frac{E}{m_e},\qquad M_e=\frac{m_e}{T},\qquad b_e=\frac{\mu_e}{T}.
\end{align}
Substituting Eq.~(\ref{ee_density}) into Eq.~(\ref{charge_neutral_cond3}) and giving the value of $X_p$, then the charge neutrality condition can be solved to determine $\mu_e/T$ as a function of $M_e$ and $T$. 
%Fig~~~~~~~~~~~~~~~~~~~~~~~~~~~~~~~~~~~~~~~~~~~~~~~~~~~~~
\begin{figure}[ht]
\begin{center}
\includegraphics[width=0.95\linewidth]{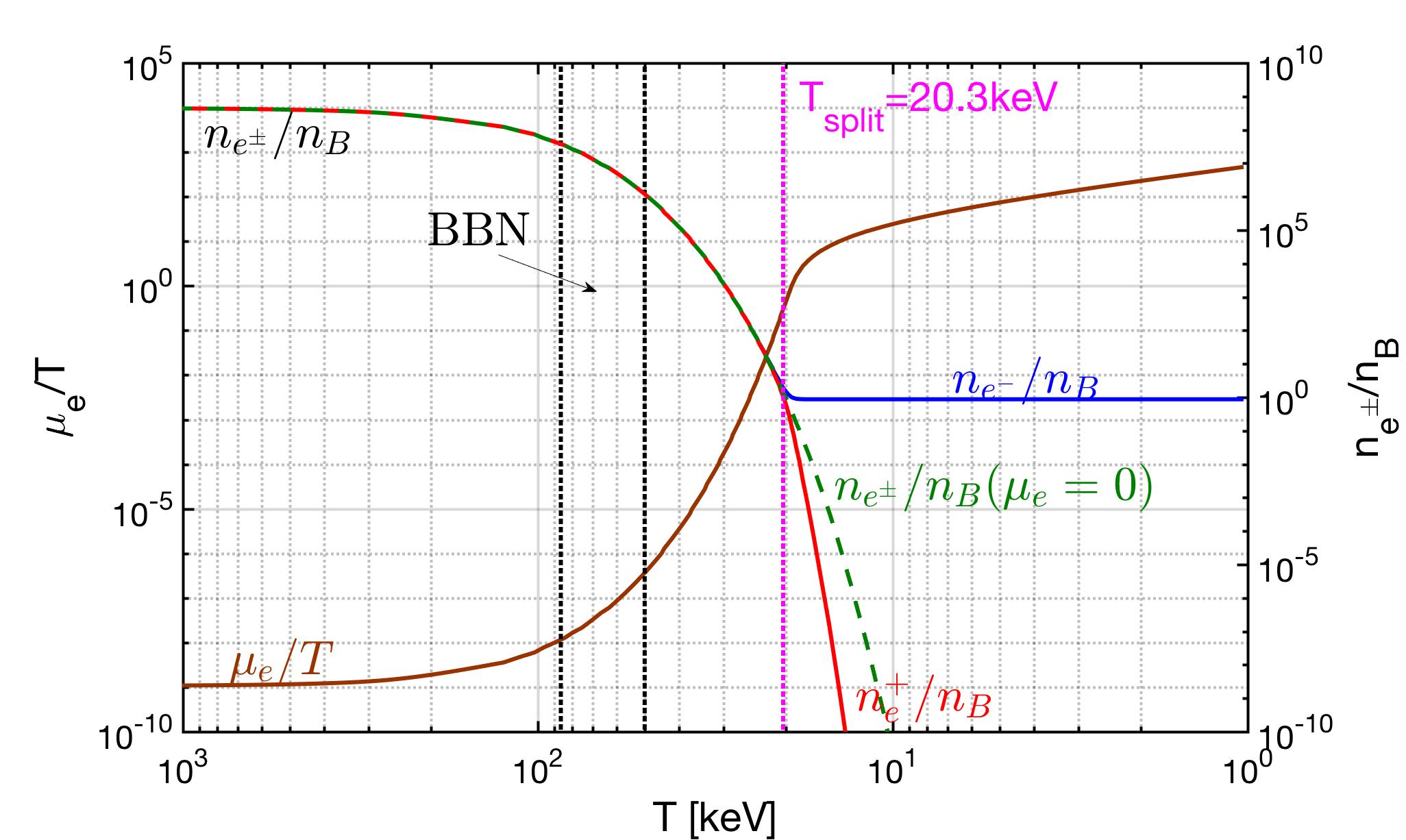}
\caption{Left axis: The chemical potential of electrons as a function of temperature by numerically solving Eq. (\ref{charge_neutral_cond3}) with $n_p/n_B=0.878$ and $n_B/n_\gamma=6.05\times10^{-10}$. Right axis: the ratio of electron (positron) number density to baryon density as a function of temperature. The solid blue line is the electron density, the red dashed line is the positron density, and the green dotted line is the number density with $\mu_e=0$. When $T=20.3\,\mathrm{keV}$ (the purple vertical line) positron density decreases rapidly because of the annihilation. The vertical black dotted lines represent BBN temperature range $86\,\mathrm{keV}>\mathrm{T_{BBN}}>50\,\mathrm{keV}$.}
\label{BBN_Electron}
\end{center}
\end{figure}
%~~~~~~~~~~~~~~~~~~~~~~~~~~~~~~~~~~~~~~~~~~~~~~~~~~~~~

In Fig.~\ref{BBN_Electron} (left axis), we solve Eq.~(\ref{charge_neutral_cond3}) numerically and plot the electron chemical potential as a function of temperature with the following parameters: proton concentration $X_p=0.878$ from observation~\cite{ParticleDataGroup:2022pth} and $n_B/n_\gamma=6.05\times10^{-10}$ from CMB. We can see the value of chemical potential is comparatively small $\mu_e/T\approx10^{-6}\sim10^{-7}$ during the BBN temperature range, implying an equal number of electrons and positrons in plasma. The ratio of electron (positron) number density to baryon density shows that the Universe was filled with an electron-positron rich plasma during the accepted BBN temperature range. For example, when the temperature is around $T=70\,\mathrm{keV}$, the density of electrons and positrons is comparatively large $n_{e^\pm}\approx10^7\,n_B$. At $90$\,keV, the electron and positron density is near the solar core density, see Fig.~19 in Ref.~\cite{Rafelski:2023emw}. Later when the temperature is around $T=20.3\,\mathrm{keV}$, the positron density decreases, transforming the pair-plasma to an electron-baryon plasma.

%%%%%%%%%%%%%%%%%%%%%%%%%%%%%%%%%%%%%%%%%%%%%%%%%%%%%%%%%%%%%%%%%%%

\section{The relaxation rate in electron-positron plasma}\label{sec:relax}
In electron-positron plasma, the major reactions between photons and $e^+e^-$ pairs are inverse Compton scattering, M{\o}ller scattering, and Bhabha scattering:
\begin{align}
&e^\pm+\gamma\longrightarrow e^\pm+\gamma,\qquad e^\pm+e^\pm\longrightarrow e^\pm+e^\pm,\qquad e^\pm+e^\mp\longrightarrow e^\pm+e^\mp.
\end{align}
In general, to evaluate the reaction rate in two-body reaction $1+2\rightarrow3+4$ in the Boltzmann approximation we can use reaction cross-section $\sigma(s)$ and the relation from Ref.~\cite{Letessier:2002gp}:
\begin{align}\label{GeneralRate}
R_{12\rightarrow34}=\frac{g_1g_2}{32\pi^4}\frac{T}{1+I_{12}}\!\!\int^\infty_{s_{th}}\!\!\!\!ds\,\sigma(s)\frac{\lambda_2(s)}{\sqrt{s}}\!K_1\!\!\left({\sqrt{s}}/{T}\right),
\end{align}
where $K_1$ is the first-order Bessel function and the K\"{a}ll\'{e}n function $\lambda_2(s)$ is defined as
\begin{align}
\lambda_2(s)=\left[s-(m_1+m_2)^2\right]\left[s-(m_1-m_2)^2\right],
\end{align}
with $m_{1/2}$ and $g_{1/2}$ are the masses and degeneracies of initial interacting particles. The factor $1/(1+I_{12})$ is introduced to avoid double counting of indistinguishable pairs of particles, we have $I_{12}=1$ for identical particles and $I_{12}=0$ otherwise. 

The two-body cross-section can be obtained by averaging the matrix element over the Mandelstam variable $t$. We have
\begin{align}
&\sigma_{e^\pm\gamma}=\frac{1}{16\pi(s-m_e^2)^2}\int^0_{-(s-m_e^2)^2/s}\!\!\!\!\!\!\!\!\!\!\!\!\!\!\!\!\! dt\, |M_{e^\pm\gamma}|^2,\\
&\sigma_{e^-e^-}=\frac{1}{16\pi s(s-4m_e^2)}\int^0_{-(s-4m_e^2)}\!\!\!\!\!\!\!\!\!\!\!\!\!\!\!\!\!dt\, |M_{e^\pm e^\pm}|^2,\\
&\sigma_{e^-e^+}=\frac{1}{16\pi s(s-4m_e^2)}\int^0_{-(s-4m_e^2)}\!\!\!\!\!\!\!\!\!\!\!\!\!\!\!\!\!dt\, |M_{e^\pm e^\mp}|^2.
\end{align}
The matrix element associated with inverse Compton scattering is given by~\cite{Kuznetsova:2009bq, Kuznetsova:2011wt}
\begin{align}
|M_{e^\pm\gamma}|^2\!&=32 \pi^2\alpha^2\bigg[4\left(\frac{m_e^2}{m_e^2-s}+\frac{m_e^2}{m_e^2-u}\right)^2\notag\\
& -\frac{4m_e^2}{m_e^2-s}-\frac{4m_e^2}{m_e^2-u} -
 \frac{m_e^2-u}{m_e^2-s} -\frac{m_e^2-s}{m_e^2-u}\bigg],
\end{align}
and for M{\o}ller and Bhabha scattering we have respectively
\begin{align}
&|M_{e^{-}e^{-}}|^{2}\!=64\pi^{2}\alpha^{2}\bigg[
\frac{s^{2}+u^{2}+8m_e^{2}(t-m_e^{2})}{2(t-m^2_{\gamma})^{2}} \notag\\ 
&+\frac{{s^{2}+t^{2}}+8m_e^{2}
(u-m_e^{2})}{2(u-m_{\gamma}^2)^{2}} + \frac{\left( {s}-2m_e^{2}\right)\left({s}-6m_e^{2}\right)}
{(t-m_{\gamma}^2)(u-m_{\gamma}^2)} \bigg],
\end{align}
and
\begin{align}
 &|M_{e^- e^+}|^{2}=64\pi^{2}\alpha^{2}
\bigg[\frac{s^{2}+u^{2}+8m_e^{2}(t-m_e^{2})}{2(t-m^2_{\gamma})^{2}}\notag\\
&+\frac{u^{2}+t^{2}+8m_e^{2}
(s-m_e^{2})}{2(s-m^2_{\gamma})^{2}} + \frac{\left({u}-2m_e^{2}\right)\left({u}-6m_e^{2}\right)}
 {(t-m^2_{\gamma})(s-m^2_{\gamma})} \bigg],
\label{M_fi_b}
\end{align}
where $s, t, u$ are the Mandelstam variables and we included the photon mass $m_\gamma$ to avoid singularity in reaction matrix elements. The photon mass in plasma is equal to the plasma frequency, in our case we have~\cite{Kislinger:1975uy}
\begin{equation}
m^2_\gamma=\omega^2_{p}=8\pi\alpha\int\frac{d^3p_e}{(2\pi)^3}\left(1-\frac{p_e^2}{3E_e^2}\right)\frac{f_- +f_+}{E_e},
\end{equation}
where $f_-$ and $f_+$ are the single-particle distribution functions for electrons and positrons respectively, and $E_e=\sqrt{p_e^2+m^2_e}$ is the electron energy. In the BBN temperature range $m_e\gg T$ and if we consider the non-relativistic electron and positron plasma
\begin{align}
m^2_\gamma=\frac{4\pi\alpha}{2m_e}\left(\frac{2m_eT}{\pi}\right)^{3/2}e^{-m_e/T}\cosh\left(\frac{\mu_e}{T}\right).
\end{align}

\begin{figure}[ht]
\begin{center}
\includegraphics[width=0.95\linewidth]{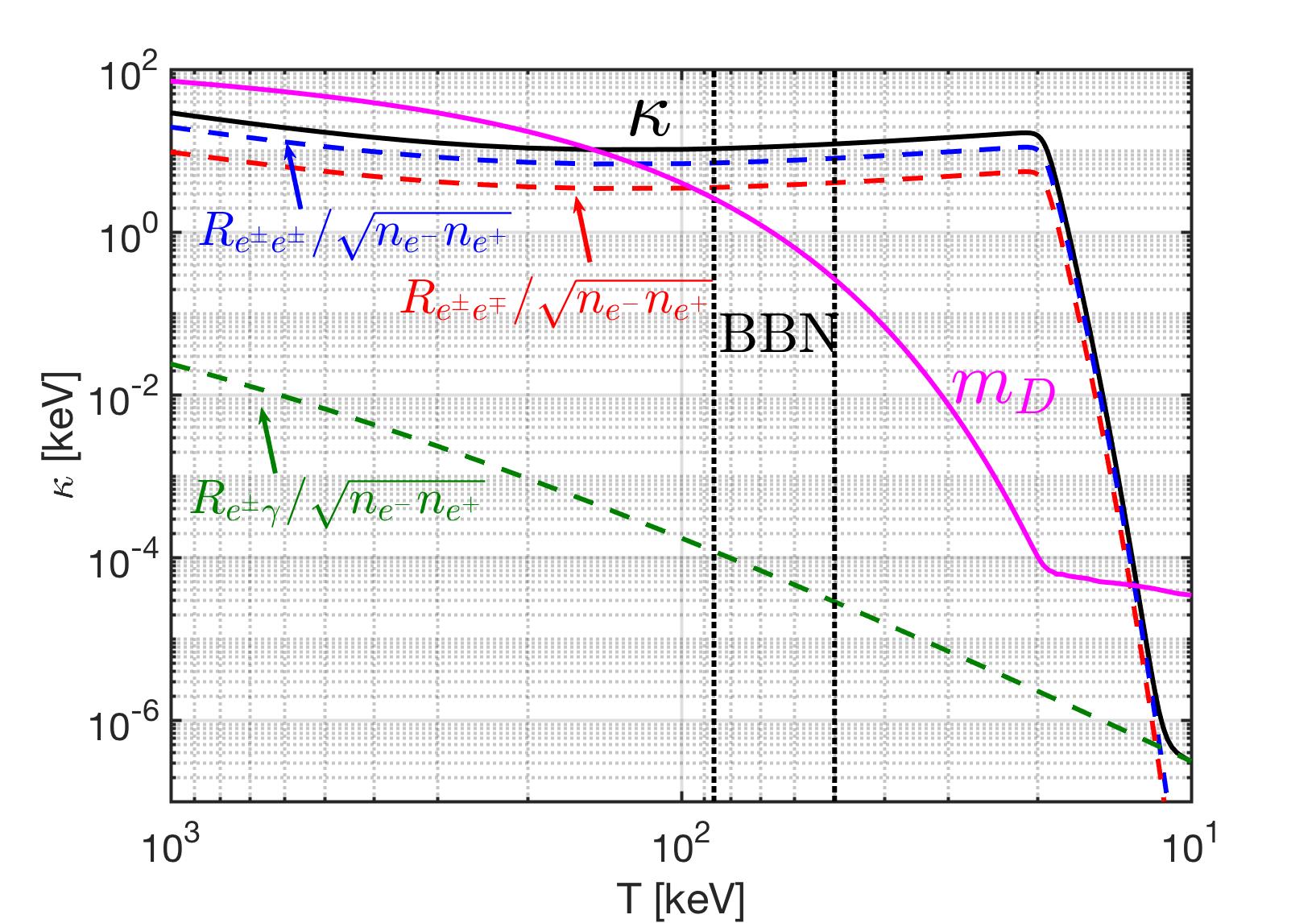}
\caption{The relaxation rate $\kappa$ as a function of temperature in non-relativistic electron-positron plasma. We present reaction rates for M{\o}ller scattering $R_{e^\pm e^\pm}$ (blue dashed line), Bhabha scattering $R_{e^\pm e^\mp}$ (red dashed line), and inverse Compton scattering $R_{e^\pm \gamma}$ (green dashed line). The average relaxation rate from Eq.~(\ref{Kappa}) is shown as a black solid line. The vertical black dotted lines represent BBN temperature range $86\,\mathrm{keV}>\mathrm{T_{BBN}}>50\,\mathrm{keV}$ during which the average relaxation rate is $\kappa=10\sim12$ keV. The dominant reactions during BBN are the M{\o}ller and Bhabha scatterings. The purple solid line represents the Debye mass given by Eq.~(\ref{eq:mL})}.
\label{RelaxationRate_fig}
\end{center}
\end{figure}

Substituting the cross-sections into Eq.~(\ref{GeneralRate}), the thermal reaction rate per volume for inverse Compton scattering can be written as
\begin{align}
R_{e^\pm\gamma}=\frac{g_eg_\gamma}{16\left(2\pi\right)^5}T\int_{m_e^2}^\infty\!\!\!\!ds\frac{K_1(\sqrt{s}/T)}{\sqrt{s}}\int^0_{-(s-m_e^2)^2/s}\!\!\!\!\!\!\!\!\!\!\!\!\!\!\!\!\!\!\!\!\!\!dt\, |M_{e^\pm\gamma}|^2,
\end{align} 
for M{\o}ller and Bhabha scattering we have
\begin{align}
&R_{e^\pm e^\pm}=\frac{g_eg_e}{16\left(2\pi\right)^5}T\!\!\int_{4m_e^2}^\infty\!\!\!\!ds\frac{K_1(\sqrt{s}/T)}{\sqrt{s}}\int^0_{-(s-4m_e^2)}\!\!\!\!\!\!\!\!\!\!\!\!\!\!\!\!\!\!\!\!\!\!dt\,|M_{e^\pm e^\pm}|^2,\\
&R_{e^\pm e^\mp}=\frac{g_eg_e}{16\left(2\pi\right)^5}T\!\!\int_{4m_e^2}^\infty\!\!\!\!ds\frac{K_1(\sqrt{s}/T)}{\sqrt{s}}\int^0_{-(s-4m_e^2)}\!\!\!\!\!\!\!\!\!\!\!\!\!\!\!\!\!\!\!\!\!\!dt\,|M_{e^\pm e^\mp}|^2.
\end{align}
In Fig.~\ref{RelaxationRate_fig}, we show the reaction rates for M{\o}ller, Bhabha, and inverse Compton scattering as a function of temperature. For temperatures $T>12.0$ keV, the dominant reactions in plasma are M{\o}ller and Bhabha scatterings between electrons and positrons. Thus, we can neglect the inverse Compton scattering in the BBN temperature range.

It is convenient to define the average relaxation rate for the electron-positron in the plasma as follows:
\begin{align}\label{Kappa}
\kappa=\frac{R_{e^\pm e^\pm}+R_{e^\pm e^\mp}+R_{e^\pm\gamma}}{\sqrt{n_{e^-}n_{e^+}}}\approx\frac{R_{e^\pm e^\pm}+R_{e^\pm e^\mp}}{\sqrt{n_{e^-}n_{e^+}}},
\end{align}
where we neglect the inverse Compton scattering during BBN as discussed. The density function ${\sqrt{n_{e^-}n_{e^+}}}$ in the Boltzmann limit is given by
\begin{align}
{\sqrt{n_{e^-}n_{e^+}}}=\frac{g_e}{2\pi^3}T^3\left(\frac{m_e}{T}\right)^2K_2(m_e/T).
\end{align}

In Fig.~\ref{RelaxationRate_fig} that the total relaxation rate $\kappa$ given by Eq.~(\ref{Kappa}) is approximately constant $\kappa=10-12$ keV during the BBN. This value is much larger than the Debye mass [see definition below in Eq.~(\ref{eq:mL})]; we expect the actual value of $\kappa$ to be smaller for a self-consistent calculation in which damping is included in the reaction rates. For $T<20.3$ keV, the relaxation rate $\kappa$ decreases rapidly because the positrons disappear. At $T=12$\,keV, the inverse Compton scattering of remaining electrons becomes the dominant scattering process. 
%%%%%%%%%%%%%%%%%%%%%%%%%%%%%%%%%%%%%%%%%%%%%%%%%%%%%%%%%%%%%%%%%%%
% %%%%%%%%%%%%%%%%%%%%%%%%%%%%%%%%%%%%%%%%%%%%%%%%%%%%%%%%%%%%%%%%%%%%%%

\section{Early Universe plasma from kinetic theory}\label{sec:kinetic_theory}
We evaluate the properties of the BBN plasma by finding the polarization tensor using the linear response relation~\cite{Starke:2014tfa} 
\begin{equation}\label{eq:linresp}
 \widetilde{j}_{\mathrm{ind}}^{\mu}(k) = {\Pi^{\mu}}_{\nu}(k) \widetilde{A}^{\nu}(k)\,.
\end{equation}
here `$k$' is the conjugate variable to the configuration space `$x$' seen below. The polarization tensor ${\Pi^{\mu}}_{\nu}(k)$ can be derived from the induced current $\widetilde{j}_{\mathrm{ind}}^{\mu}(k)$ due to perturbations $\delta f$ away from equilibrium in the relativistic Vlasov-Boltzmann transport equations
\begin{align}\label{eq:VBEf}
(p \cdot \partial) f_\pm(x,p) + &q F^{\mu\nu} p_\nu \frac{\partial f_\pm(x,p)}{\partial p^\mu} = C_\pm(x,p)\,,\\
\label{eq:VBEg}(p \cdot \partial) f_\gamma(x,p) &= C_\gamma(x,p)\,.
\end{align}

The subscripts $-$, $+$, and $\gamma$ indicate the transport equation for electrons, positrons, and photons. These form a system of differential equations for each distribution function $f(x,p)$. We take the collision term $C(x,p)$ in the BGK form as given by~\cite{Bhatnagar:1954zz, Formanek:2021blc}
\begin{equation}\label{eq:collision}
 C_i (x,p) =\kappa_i (p\cdot u)\left(f_i^{\text{eq}} (p)\frac{n_i(x)}{{n_i^{\text{eq}}}} - f_i(x,p)\right)\,,
\end{equation}
where $u$ is the 4-velocity of the plasma, $\kappa$ is the damping coefficient, and $n$ is plasma particle the density. The subscripts "i" indicate the distributions of fermions and photons, respectively. We suppressed the 4-momentum subscript for each species $f_i(x,p) = f_i(x,p_i)$ to simplify notation. Plasma constituents collide on a momentum-averaged time scale $\tau_{\text{rel}} = \kappa^{-1}$, or the "mean free time" in the plasma. The effect of the BGK collision term Eq.\,(\ref{eq:collision}) is to model the sum of all scattering effects on a species as a dissipative medium effect which returns the system to equilibrium. The BGK collision term is constructed such that Eq.\,(\ref{eq:VBEf}) retains current conservation~\cite{Bhatnagar:1954zz}.

%%%%%%%%%%%%%%%%%%%%%%%%%%%%%%%%%%%%%%%%%%%%%%%%%%%%%%%%%%%%%%%%%%%

 %%%%%%%%%%%%%%%%%%%%%%%%%%%%%%%%%%%%%%%%%%%%%%%%%%%%%%%%%%%%%%%%%%%
%%%%%%%%%%%%%%%%%%%%%%%%%%%%%%%%%%%%%%%%%%%%%%%%%%%%%%%%%%%%%%%%%%%%%%
Since photons cannot couple directly to the electromagnetic field, they do not contribute to the polarization tensor at first order in $\delta f$ as indicated in Eq.\,(\ref{eq:VBEg}). We neglect photon influence on the electron-positron distribution through the scattering term since the rate of inverse Compton scattering $R_{e^{\pm}\gamma }$ shown in green in \reff{RelaxationRate_fig} is much smaller than the total rate $\kappa$ shown black in the BBN temperature range and in general when noticeable positron abundance is present. Each fermion Boltzmann equation \req{eq:VBEf} can be solved independently. Since the equations for electrons and positrons are equivalent, except for the sign of the charge, only one needs to be solved to understand the dynamics.

\begin{figure}[ht]
\begin{center}
\includegraphics[width=0.95\linewidth]{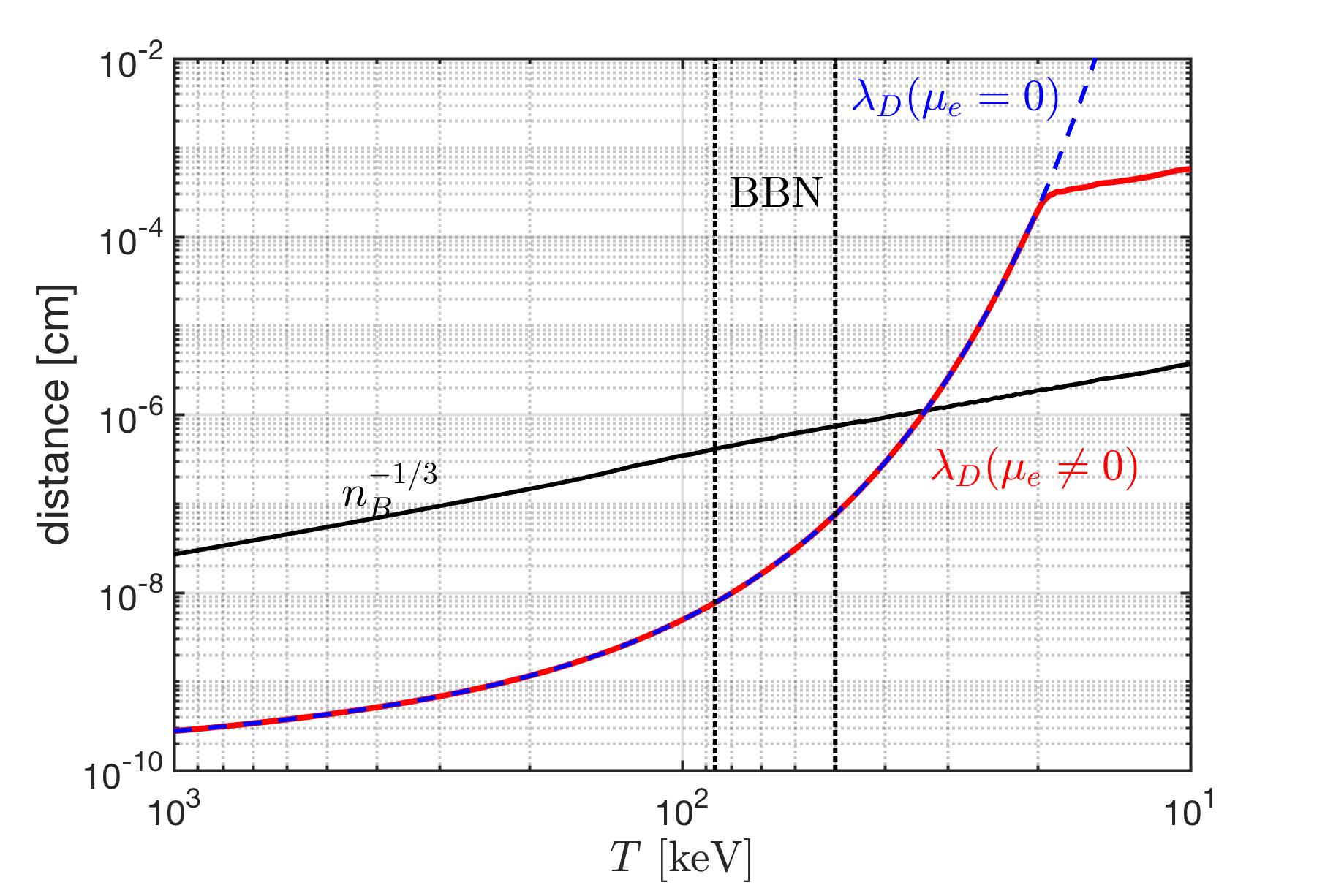}
\caption{The average distance between baryons $n_B^{-1/3}$ and the Debye length $\lambda_D$ ($\mu_e \neq 0$) as a function of temperature (red solid line). During the BBN epoch (vertical dotted lines) $n_B^{-1/3}>\lambda_D$. For temperature below $T<32.76$ keV we have $n_B^{-1/3}<\lambda_D$. For comparison, the Debye length for zero chemical potential $\mu_e=0$ is also plotted as a blue dashed line.}
\label{MeanFreePath_fig}
\end{center}
\end{figure}

We do not consider the influence of light nuclei on the polarization tensor since their density during the BBN epoch is much smaller than that of electrons and positrons \reff{BBN_Electron}. One can see in \reff{MeanFreePath_fig} that the separation of baryons $n_B^{-1/3}$ in black is much larger than the size of the polarizing Debye sphere, so baryons do not participate significantly in screening.

We take the equilibrium one particle distribution function $\eq{f_\pm}$ of electrons and positrons to be the relativistic Fermi-distribution
\begin{equation}\label{eq:equildist}
\eq{f}_\pm(p) = \frac{1}{\exp{\left(\frac{\sqrt{\boldsymbol{p}^2 + m^2}}{T}\right)}
+1}\,,
\end{equation}
with chemical potential $\mu = 0 $. The electron and positron mass will be indicated by $m$ unless otherwise stated. At temperatures interesting for nucleosynthesis $T = 50-86$\,keV, we expect the plasma temperature to be much less than the mass of the plasma constituents. Only the non-relativistic form of Eq.\,(\ref{eq:equildist}) will be relevant at these temperature scales
\begin{equation}
\eq{f}_\pm(p) \approx \exp\left(- \frac{m}{T}\left(1+\frac{|\pmb{p}|^2}{2m^2}\right)\right)\,.
\end{equation}
Keeping terms up to quadratic order in $|\boldsymbol{p}|/m$ we solve the Vlasov-Boltzmann equation [Eq.\,(\ref{eq:VBEf})] for the induced current and identify the polarization tensor. This is done in detail in our previous work in~\cite{Formanek:2021blc}, but we will reiterate a few key steps.

First, we expand Eq.\,(\ref{eq:VBEf})
around small perturbations from equilibrium
\begin{equation}\label{eq:perturbation0}
f_\pm(x,p) = {\eq{f}_\pm}(p) + \delta f_\pm(x,p)\,,
\end{equation}
and solve \req{eq:VBEf} for $\delta f_\pm(x,p)$ in Fourier space. The induced current in Fourier space is given by
\begin{equation}\label{eq:perturbation1}
\tilde{j}_{\mathrm{ind}}^\mu(k) = 2\int \frac{d^4 p}{(2 \pi)^4}p^\mu 4\pi \delta_+(p^2-m^2) 
\sum_{i = \, +, \, -} q_i \tilde{f}_{i}(k,p)\,,
\end{equation}
with the factor of two accounting for spin.
After inserting \req{eq:perturbation0}, and specifying $q_\pm = \pm e$ the induced current is a function of the perturbation
\begin{multline}\label{eq:perturbation2}
\tilde{j}_{\mathrm{ind}}^\mu(k) = 2\int \frac{d^3 p}{(2 \pi)^3 p^0}p^\mu \Big( e \left[\eq{\tilde{f}}_+(k,p)-\eq{\tilde{f}}_-(k,p)\right]\\
+ e\left[\delta\tilde{f}_+(k,p)-\delta\tilde{f}_-(k,p)\right]
\Big)
\\
=4 e\int \frac{d^3 p}{(2 \pi)^3 p^0}p^\mu \delta\tilde{f}(k,p)
\,,
\end{multline}
because the equilibrium currents cancel in the weak field limit and the perturbations add since they differ by the charge $\delta f_\pm=\pm e \delta f' $. This term is studied for finite chemical potential in~\cite{Wang:2010px}. We focus on the second term related to the polarization response of the plasma.

The polarization tensor can be obtained from \req{eq:perturbation2} by computing the 4-momentum integrals over $p$ in the rest frame of the plasma. Once integrated, the 4-potential $\widetilde{A}^{\nu}(k)$, coming from the electromagnetic field strength tensor $F^{\mu \nu}$ in \req{eq:VBEf}, factors out since it only depends on the 4-wavevector $k$ and \req{eq:perturbation2} attains the form of \req{eq:linresp}. For details, see Ref.~\cite{Formanek:2021blc}.

In the infinite homogeneous plasma filling the early Universe, the polarization tensor only has two independent components: the longitudinal polarization function $\Pi_{\parallel}$ parallel to field wave-vector $\boldsymbol{k}$ in the rest frame of the plasma and the transverse polarization function $\Pi_{\perp}$ perpendicular to $\boldsymbol{k}$~\cite{Melrose:2008}. In the non-relativistic limit, these functions are~\cite{Formanek:2021blc}
\begin{align}\label{eq:polfuncs}
	\Pi_\parallel(\omega,\boldsymbol{k}) &= -\omega_p^2\frac{\omega^2}{(\omega+ i \kappa)^2} \frac{1}{1-\frac{i\kappa}{\omega+ i \kappa}\left(1+\frac{ T |\boldsymbol{k}|^2}{m (\omega+ i \kappa)^2} \right)}\,,\\
	\Pi_{\perp}(\omega) &= -\omega_p^2 \frac{\omega}{\omega+ i \kappa}\,.
\end{align}
In these expressions, the plasma frequency $\omega_p$ (defined as $m_L$ in Ref.~\cite{Formanek:2021blc}) is related to the Debye screening mass in the non-relativistic limit as
\begin{equation}\label{eq:plasmafreq}
 \omega_p^2 = m_D^2\frac{T}{m}\,.
\end{equation}

%~~~~~~~~~~~~~~~~~~~~~~~~~~~~~~~~~~~~~~~~~~~~~

The transverse response $\Pi_{\perp}$ relates to the dispersion of photons in the plasma. Here we need only consider $\Pi_\parallel$ since the vector potential $\boldsymbol{A}(t,\boldsymbol{x})$ of the traveling ion will be small in the non-relativistic limit. In this work, we do not consider the effect of a primordial magnetic field discussed in Ref.~\cite{Steinmetz:2023abc}. We note that Debye mass $m_D$ is related to the usual Debye screening length of the field in the plasma as
\begin{equation}\label{eq:mL}
	1/\lambda_D^{2} = m_D^2= 4 \pi \alpha \left(\frac{2mT}{\pi}\right)^{3/2}\frac{e^{-m/T}}{2T}\,.
\end{equation}
This formula describes the characteristic length scale of screening in the plasma.

\section{Effective internuclear potential}\label{sec:potential}
We calculate the potential of light nuclei in the early Universe electron-positron plasma by finding the screened scalar potential $\phi$ of a single traveling nuclei
\begin{equation}\label{eq:potent}
 \phi(t,\boldsymbol{x}) = \int \frac{d^4k}{(2\pi)^4} e^{-i\omega t+ i\boldsymbol{k}\cdot\boldsymbol{x}} \frac{\widetilde{\rho}_\text{ext}(\omega,\boldsymbol{k})}{\varepsilon_\parallel(\omega,\boldsymbol{k})(\boldsymbol{k}^2-\omega^2) }\,,
\end{equation}
where $\widetilde{\rho}_{\text{ext}}(\omega,\boldsymbol{k})$ is the Fourier-transformed charge distribution of nuclei traveling at a constant velocity, as found in \ref{sec:freechg}, and $\varepsilon_\parallel(\omega,\boldsymbol{k})$ is the longitudinal relative permittivity.\footnote{We omit $\varepsilon_0$ until the final result, to avoid confusion with $\varepsilon_\parallel(\omega,\boldsymbol{k})$.} The screened potential is obtained by solving Maxwell equations algebraically in Fourier space in the presence of a polarizing medium. The relative permittivity can be written in terms of the polarization tensor as
\begin{equation}\label{eq:epsilon}
 \varepsilon_\parallel(\omega,\boldsymbol{k})= \left(\frac{\Pi_{\parallel}(\omega,\boldsymbol{k})}{ \omega^2}+1\right)\,.
\end{equation}

The potential between two nuclei can be inferred simply from the potential of a single nucleus, since in the linear response framework \req{eq:linresp} the electromagnetic field still obeys the principle of superposition. 

We can perform the $\omega$ integration in \req{eq:potent} using the delta function in the definition of the external charge distribution \req{eq:extchgfreq}, whose effect is to set $\omega = \boldsymbol{\beta_{\text{N}}}\cdot \boldsymbol{k}$ where $\boldsymbol{\beta}_N = \boldsymbol{v}_N/c$ is the nuclei velocity. Then we have
\begin{equation}\label{eq:potentk}
 \phi(t,\boldsymbol{x}) = Ze\int \frac{d^3\boldsymbol{k}}{(2\pi)^3} e^{ i\boldsymbol{k}\cdot(\boldsymbol{x}-\boldsymbol{\beta_{\text{N}}} t)} \frac{ e^{-\boldsymbol{k}^2\frac{R^2}{4}}}{\boldsymbol{k}^2\varepsilon_\parallel(-\boldsymbol{\beta_{\text{N}}} \cdot \boldsymbol{k},\boldsymbol{k}) }\,,
\end{equation}
where $R$ is the Gaussian radius parameter.
In non-relativistic approximation the Lorentz factor $\gamma \approx 1$ and we use the convention $\varepsilon_\parallel(-\boldsymbol{\beta_{\text{N}}} \cdot \boldsymbol{k},\boldsymbol{k})$~\cite{Montgomery:1970jpp, Stenflo:1973,Shukla:2002ppcf,Shukla1996} which gives the correct causality for the potential. Specifically, without damping, the wakefield occurs behind the moving nucleus.

In the limit of a single nucleus at rest in the plasma, the screened potential has the usual Debye-H\"uckel form~\cite{Debye:1923}
\begin{equation}
 \phi_{\text{stat}}(t,\boldsymbol{x}) = \frac{Z e}{4 \pi \varepsilon_0}\frac{e^{-m_D r}}{r}\,.
\end{equation}
The analogous formula for the static screening of a Gaussian charge is derived in \ref{sec:static}.
%%%%%%%%%%%%%%%%%%%%%%%%%%%%%%%%%%%%%%%%%%%%%%%%%%%%%%%%%%%%%%%%%%%%%%%%%%%%%%%
\subsection{Damped-dynamic screening}\label{sec:DDS}
In this section, we compute the effective nuclear potential for a light nucleus moving in the plasma at a constant velocity. This is done by Fourier transforming \req{eq:potentk}. The velocity of the nucleus is assumed to be the most probable velocity given by a Boltzmann distribution
\begin{equation}
 \beta_{\text{N}} = \sqrt{\frac{2T}{m_N}}\,. 
\end{equation}
Since the poles of the \req{eq:potent} can be solved analytically, ideally, one would perform contour integration to get the position space field. Due to the intricacy of these poles $\omega_n(\boldsymbol{k})$, we find it insightful to look at the field in a series expansion around velocities of the light nuclei smaller than the thermal velocity of electrons and positrons, and large damping.

% \begin{equation}
% \beta_{\text{N}}\frac{m_D}{\kappa} = \frac{v_{\text{N}}}{c}\frac{m_D}{\kappa} \ll 1\,.
% \end{equation}
\begin{equation}\label{eq:expansion}
% (\boldsymbol{k}\cdot\boldsymbol{\beta}_{\text{N}})^2 \ll \boldsymbol{k}^2 \frac{T}{m} \ll \kappa^2\, 
\frac{(\boldsymbol{k}\cdot\boldsymbol{\beta}_{\text{N}})^2}{\omega_p^2} \ll \frac{\boldsymbol{k}^2}{m_D^2} \ll \frac{\kappa^2}{\omega_p^2}\, .
\end{equation}

This expansion is useful during BBN since the temperature is much lower than the mass of light nuclei and the damping rate $\kappa$ is approximately twice the Debye mass $m_D$, as seen in Fig.~\ref{RelaxationRate_fig}. Applying this expansion to \req{eq:potentk} we obtain at first order the potential
% \begin{multline} \label{eq:potexp}
% \phi(t,\boldsymbol{x}) = \phi_{\text{stat}}(t,\boldsymbol{x}) +\\ -Ze\int \frac{d^3\boldsymbol{k}}{(2\pi)^3} e^{ i\boldsymbol{k}\cdot(\boldsymbol{x}-\boldsymbol{\beta_{\text{N}}} t)}\frac{i \boldsymbol{k}\cdot \boldsymbol{\beta_{\text{N}}} m_D^2 (\frac{\boldsymbol{k}^2}{\kappa} - \frac{m}{T} \kappa)}{\boldsymbol{k}^2(\boldsymbol{k}^2+m_D^2)^2} e^{-\boldsymbol{k}^2\frac{R^2}{4}} \\ + O
% \left(\beta_{\text{N}}\frac{m_D}{\kappa}^2\right)\,.
% \end{multline}
\begin{multline} \label{eq:potexp}
 \phi(t,\boldsymbol{x}) = \phi_{\text{stat}}(t,\boldsymbol{x})\\ -Ze\int \frac{d^3\boldsymbol{k}}{(2\pi)^3} e^{ i\boldsymbol{k}\cdot(\boldsymbol{x}-\boldsymbol{\beta_{\text{N}}} t)}\frac{i \boldsymbol{k}\cdot \boldsymbol{\beta_{\text{N}}} m_D^4 (\frac{\boldsymbol{k}^2}{m_D^2} - \frac{\kappa^2}{\omega_p^2})}{\boldsymbol{k}^2(\boldsymbol{k}^2+m_D^2)^2\kappa} e^{-\boldsymbol{k}^2\frac{R^2}{4}} \,.
\end{multline}
First, we focus on the second term evaluating for a point charge $R\rightarrow 0$
\begin{equation}
 \Delta\phi(t,\boldsymbol{x}) =-Ze\int \frac{d^3\boldsymbol{k}}{(2\pi)^3} e^{ i\boldsymbol{k}\cdot(\boldsymbol{x}-\boldsymbol{\beta_{\text{N}}} t)}\frac{i \boldsymbol{k}\cdot \boldsymbol{\beta_{\text{N}}} m_D^4 (\frac{\boldsymbol{k}^2}{m_D^2} - \frac{\kappa^2}{\omega_p^2})}{\boldsymbol{k}^2(\boldsymbol{k}^2+m_D^2)^2\kappa}\,.
\end{equation}
Numerical calculations are performed using the full Gaussian charge distribution. We focus on the second term in \req{eq:potexp} since $\phi_{\text{stat}}$ is the standard static screening result. In order to perform this integration we first note that this expression can be re-written as the laboratory time derivative
\begin{equation}
 \Delta\phi(t,\boldsymbol{x}) = Ze\frac{d}{dt}\left[\int \frac{d^3\boldsymbol{k}}{(2\pi)^3} e^{ i|\boldsymbol{k}||\boldsymbol{x}-\boldsymbol{\beta_{\text{N}}} t| \cos(\theta)}\frac{ m_D^4 (\frac{\boldsymbol{k}^2}{m_D^2} - \frac{\kappa^2}{\omega_p^2})}{\boldsymbol{k}^2(\boldsymbol{k}^2+m_D^2)^2\kappa}\right] \,,
\end{equation}
where the dot products were replaced by their angular dependence.
The angular integration can be evaluated in spherical coordinates, see \ref{sec:static} for details,
\begin{equation}
 \Delta\phi(t,\boldsymbol{x}) = 2 Ze\frac{d}{dt}\left[\int \frac{d\boldsymbol{k}}{(2\pi)^2} \frac{\sin(|\boldsymbol{k}||\boldsymbol{x}-\boldsymbol{\beta}_N t|)}{|\boldsymbol{k}||\boldsymbol{x}-\boldsymbol{\beta}_N t|}\,\frac{ m_D^4 (\frac{\boldsymbol{k}^2}{m_D^2} - \frac{\kappa^2}{\omega_p^2})}{\boldsymbol{k}^2(\boldsymbol{k}^2+m_D^2)^2\kappa}\right] \,,
\end{equation}
finally, this can be integrated over the radial wavevector to find
\begin{multline}
 \Delta\phi(t,\boldsymbol{x}) = \frac{Ze }{4\pi }\frac{m_D^2}{\kappa}\frac{d}{dt}\Bigg[\left(\frac{1+\nu_\tau^2}{ 2m_D}+\frac{\nu_\tau^2}{m_D^2 |\boldsymbol{x}-\boldsymbol{\beta}_N t|}\right)e^{-m_D |\boldsymbol{x}-\boldsymbol{\beta}_N t|}\\ - \frac{\nu_\tau^2}{m_D^2 |\boldsymbol{x}-\boldsymbol{\beta}_N t|} \Bigg]\,,
\end{multline}
here we introduce the ratio of the damping rate to the rate of oscillations in the plasma $\nu_\tau = \kappa/\omega_p$. We can apply the time derivative; note that
\begin{equation}\label{eq:der_cos}
 \frac{d}{dt} |\boldsymbol{x}-\boldsymbol{\beta}_N t| = - \frac{\boldsymbol{\beta}_N \cdot (\boldsymbol{x}-\boldsymbol{\beta}_N t)}{|\boldsymbol{x}-\boldsymbol{\beta}_N t|}= -\beta_N \cos(\psi)\,,
\end{equation}
where $\psi$ is the angle between $\boldsymbol{x}-\boldsymbol{\beta}_N t$ and $\boldsymbol{\beta}_N$. After taking the derivative and using the above expression \req{eq:der_cos} one finds
\begin{multline}\label{eq:pos_point_DDS}
\Delta \phi(t,\boldsymbol{x}) = \frac{Ze \beta_N \cos (\psi) m_D^2}{4 \pi \varepsilon_0 \kappa} \Bigg[\left(\frac{\nu_\tau^2}{m_D^2 r(t)^2} + \frac{\nu_\tau^2}{m_D r(t)}+\frac{1 + \nu_\tau^2}{2}\right)e^{-m_D r(t)} \\ -\frac{\nu_\tau^2}{m_D^2 r(t)^2}\Bigg]\,,
\end{multline}
with $r(t) = |\boldsymbol{x}-\boldsymbol{\beta}_N t|$. This expression is valid for large damping and slow motion of the nucleus or if the velocity of the nuclei is small. A similar result valid at large distances, which only includes the last term, was derived previously in Ref.~\cite{Stenflo:1973} in the context of dusty (complex) plasmas. The static screening potential $\phi_{\text{stat}}$ must be added to this contribution to get the full potential
\begin{equation}\label{eq:pos_point}
\phi(t,\boldsymbol{x}) = \phi_{\text{stat}}(t,\boldsymbol{x}) +\Delta \phi(t,\boldsymbol{x}) \,.
\end{equation}
For large distances and large $\nu_\tau$, the last term in the second line is dominant, indicating that the overall potential would be over-damped. In this regime, the potential is heavily screened in the forward direction and unscreened in the backward direction relative to the motion of the nucleus. As $\nu_\tau$ becomes small, the $1/2$ in the first portion of the third term, proportional to $m_D^2/\kappa$, dominates. This flips the sign of the damped-dynamic screening contribution causing a wake potential to form behind the nuclei. This shift indicates the change from damped to undamped screening where \req{eq:pos_point_DDS} is no longer valid.

We note in passing that using a Schwinger parameterization similar to the one used to derive \req{eq:schwin}
\begin{equation}
\frac{1}{(k^2+m_D^2)^2} =\int_0^\infty ds \, s \, e^{-s(k^2 +m_D^2)}\,,
\end{equation} 
it is possible to derive an expression for damped-dynamic screening for a Gaussian charge distribution, which is a trivial extension of \req{eq:pos_point_DDS}.

\begin{figure}[ht]
 \centering
 \includegraphics[width=.95\linewidth]{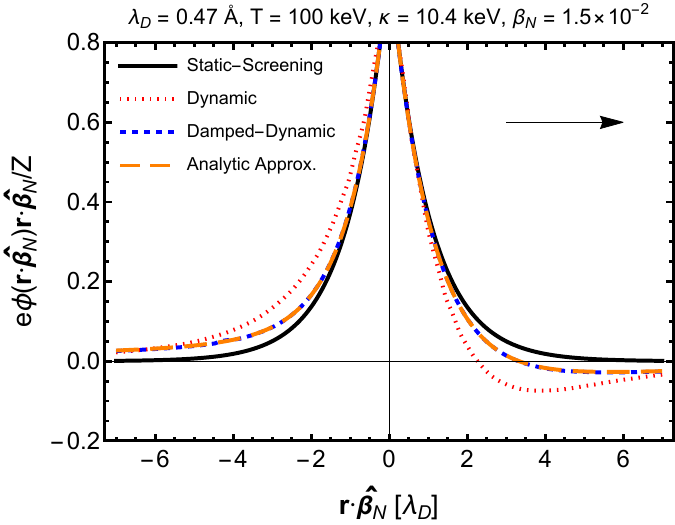}
 \caption{A comparison of the following screening models plotted along the direction of motion of a nucleus $\boldsymbol{r}\cdot\hat{\boldsymbol{\beta}_{\text{N}}}$: static screening (black), dynamic screening (red dotted)\cite{Hwang:2021kno}, damped-dynamic screening (blue dashed), and the approximate analytic solution of \req{eq:pos_point} (orange dashed). A black arrow indicates the direction of motion of the nucleus $\hat{\boldsymbol{\beta}_{\text{N}}}$. }
 \label{fig:dynamiclinear}
\end{figure} 

Figure \ref{fig:dynamiclinear} demonstrates that the damped-dynamic response in the analytic approximation \req{eq:pos_point_DDS} (shown as orange dashed line) is sufficient to approximate the full numerical solution (blue dashed line) found by numerical integration of \req{eq:potent}. The temperature $T = 100$\, keV, above our upper limit of BBN temperatures, is chosen to relate to the dynamic screening result found in Ref.~\cite{Hwang:2021kno}. Our analytic solution differs from the numerical result in Fig.~4 of Ref. \cite{Hwang:2021kno} by a factor of $\sqrt{2}$ and is horizontally flipped. This reflection is due to a difference in convention in the permittivity, as seen in \req{eq:potentk}. We can see that at large distances, dynamic screening is slightly stronger than damped screening, as expected. At short distances relevant for thermonuclear reaction rates, damped and undamped screening are very similar. 

Dynamic screening in both the damped and undamped cases predicts less screening behind and more in front of the moving nucleus, as compared to static screening. This effect was previously observed for subsonic screening in electron-ion-dust plasmas ~\cite{Stenflo:1973, Shukla:2002ppcf, Lampe:2000pop}. As a result, a negative polarization charge builds up in front of the nucleus. The small negative potential in front alters the potential energy between light nuclei, possibly changing the equilibrium distribution of light elements. This effect is much larger in the undamped case and is known in some cases to lead to the formation of dust crystals~\cite{Shukla1996}. We discuss possible implications for the early universe structure formation in the conclusions.

\begin{figure}[ht]
 \centering
 \includegraphics[width=.95\linewidth]{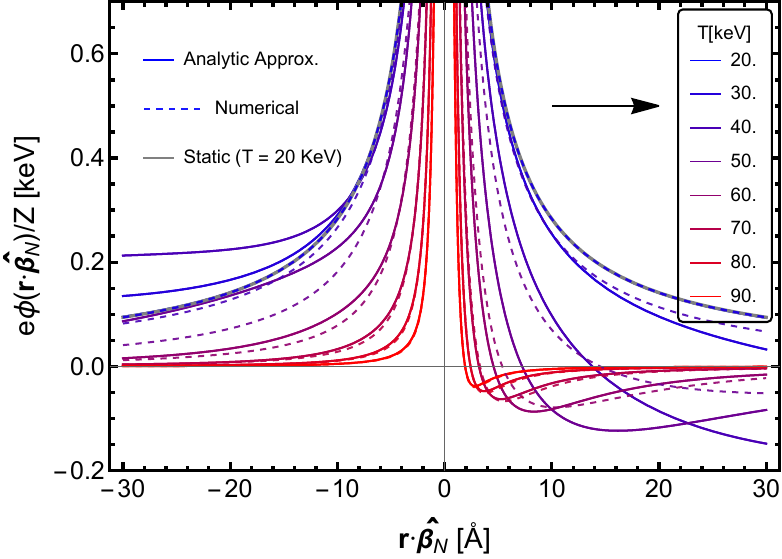}
 \caption{The analytic result from \req{eq:pos_point_DDS} is compared to the full numerical (dashed lines) result plotted along the direction of motion of a nucleus in the temperature range, $20 - 90 \,$keV. At low temperatures, the full numerical solution returns to static screening (gray solid line). At high temperatures near the beginning of BBN, the \req{eq:pos_point_DDS} approaches the full numerical solution.}
 \label{fig:numericalComp}
\end{figure} 

In \reff{fig:numericalComp}, we present a comparison of the approximate analytic result of \req{eq:pos_point_DDS}, shown as solid lines ranging from blue to red, to the full numerical solution of \req{eq:potent}, shown as dashed lines, for damped-dynamic screening. Interestingly, as the temperature decreases, the screening effect increases in contrast to the usual Debye-H\"uckel screening behavior. This is because in the large damping limit, we have terms in \req{eq:pos_point_DDS} proportional to $\kappa/m_D^2$. Since the debye mass decreases with temperature but $\kappa$ remains relatively constant during BBN \reff{RelaxationRate_fig}, one finds an overall increase in long-distance screening. The expansion used to derive the analytic solution \req{eq:expansion} assumes large temperatures and breaks down at temperatures below $60\,$keV. At low temperatures, the numerical result matches the static screening result; at $T=20\,$kev, the numerical solution indicated by a dashed blue line traces static screening at the same temperature, shown in solid gray. The analytic solution trends to the numerical result at higher temperatures, within $\sim 10\%$ for $T = 70 \,$keV. At temperatures above $200\,$keV, the non-relativistic approximation should be re-evaluated.

\subsection{Limitations of linear response: toy model}
\label{ssec:limitLR}
We now consider the equilibrium phase space distribution in the presence of a `strong' potential. In covariant form and not neglecting the magnitude of the 4-potential $A^\mu$ as compared to the energy-momentum $p^\mu$ of the particle in statistical ensemble, one obtains~\cite{Hakim:1967prd,DeGroot:1980dk}
\begin{equation}
 \eq{f}(x,p) = e^{-u_{\mu}(p^{\mu}+q A_{(\text{eq})}^{\mu}(x))/T}\,,
\end{equation}
where $u^{\mu} = (1,0,0,0)$ is the global velocity of the plasma in its rest frame. The density according to this distribution is given by the usual expression but altered by the exponential factor in potential
\begin{equation}
 n_{eq} = \int \frac{dp^3}{(2 \pi)^3 p^0} p^0 \eq{f}(x,p) = n_{eq} e^{-u^{\mu}q A_{(\text{eq})}^{\mu}(x)/T}\,.
\end{equation}
Inserting the equilibrium density into the Poisson equation
\begin{equation}
 -\nabla^2 \phi_{(\text{eq})}(x) =\rho_\mathrm{tot}(x)=\rho_\mathrm{ext}(x)+\rho_\mathrm{ind}(x)\,,
\end{equation}
gives the nonlinear Poisson-Boltzmann equation in equilibrium
\begin{equation}
 -\nabla^2 \phi_{(\text{eq})}(x) +4e n_{(\text{eq})} \sinh\left[e\phi_{(\text{eq})}(x)/T\right] =\rho_\mathrm{ext}(x)\,.
\end{equation}
We can re-scale the potential with temperature to rewrite as
\begin{equation}\label{eq:Poisson-Boltz}
 -\nabla^2 \Phi_{(\text{eq})}(x) +m_D^2\sinh\left[\Phi_{(\text{eq})}(x)\right] =e\rho_\mathrm{ext}(x)/T\,,
\end{equation}
where $\Phi_{(\text{eq})}(x) = e\phi_{(\text{eq})}(x)/T$. This equation has a well-known solution for an infinite sheet, where the external charge $\rho_\mathrm{ext}$ becomes a fixed boundary condition. We will set the potential at the sheet $\phi(0) = \phi_0$ and solve the Poisson-Boltzmann equation in the region to the right of the sheet ($x > 0$)
\begin{equation}
 -\frac{d^2}{dx^2}\Phi_{(\text{eq})}(x) +m_D^2\sinh\left[\Phi_{(\text{eq})}(x)\right] =0\,.
\end{equation}
The solution to this problem can be found for example in Ref.~\cite{Gouy:1910}
\begin{equation}\label{eq:nonlinscreen}
 \Phi_{(\text{eq})}(x) = \frac{e\phi_{(\text{eq})}(x)}{T} = 4 \tanh^{-1} \left[\tanh\left(\frac{e\phi_0}{4T}\right)e^{- m_D x}\right]\,,
\end{equation}
which we can compare to the usual Debye screening for a charged plane
\begin{equation}\label{eq:linscreen}
 \phi_{(\text{eq})}(x) =\phi_0e^{-m_D x}\,.
\end{equation}

\begin{figure}[ht]
 \centering
 \includegraphics[width=.95\linewidth]{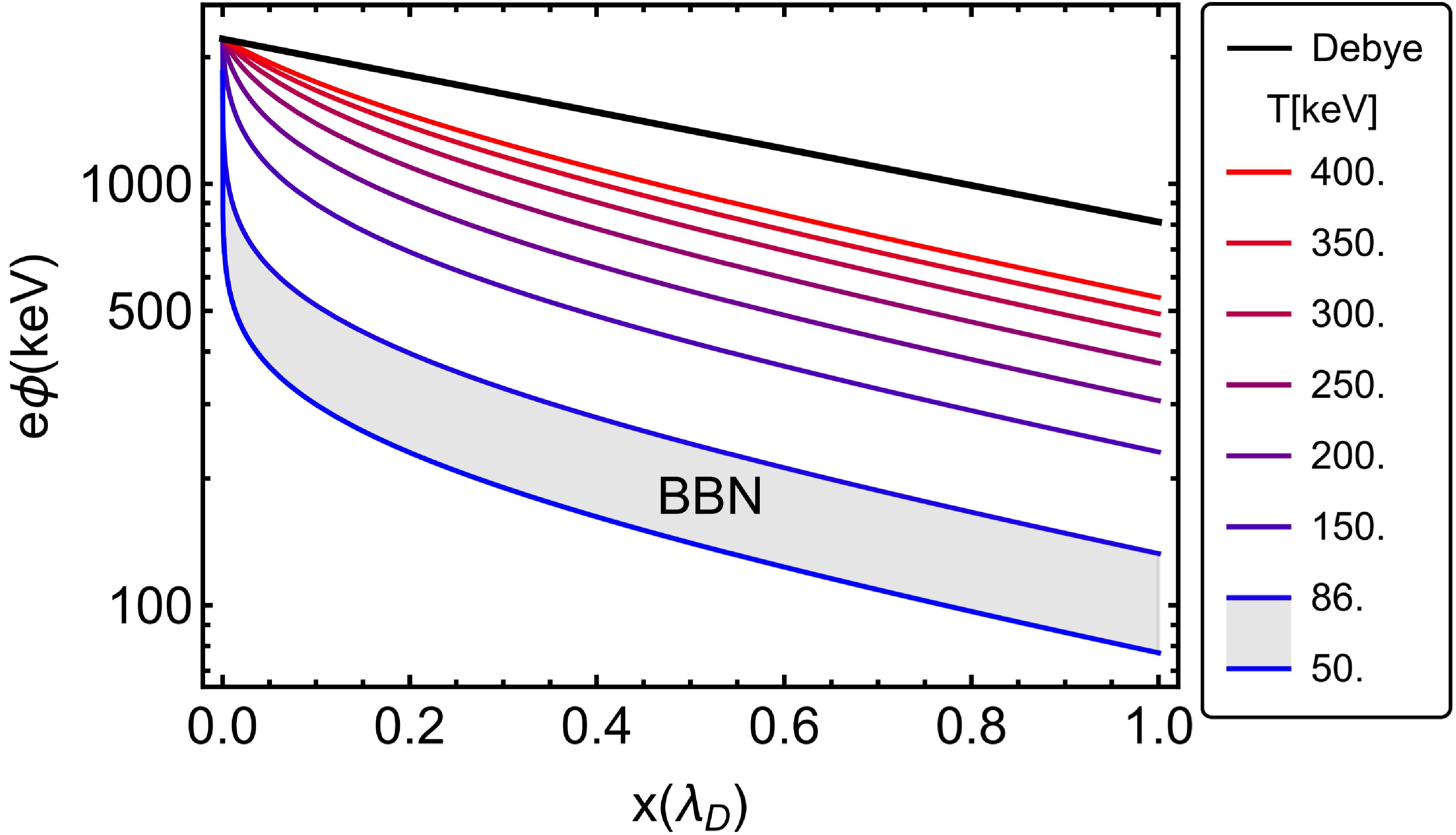}
 \caption{Comparison of nonlinear screening to Debye linearized screening (solid top line) in the range $50\le T\le 400$\,keV; the BBN temperature range is shaded. In this toy model example $Z=2$ was used in \req{eq:phi0}.
 } \label{fig:nonlinscreen}
\end{figure}

We contrast linear screening \req{eq:linscreen} and nonlinear screening \req{eq:nonlinscreen} in \reff{fig:nonlinscreen}. The boundary value of the potential $\phi_0$ is chosen to be the maximum value of the potential of a Gaussian charge distribution
\begin{equation}\label{eq:phi0}
 \phi_0 = \frac{Z e}{8 \pi^{3/2} \varepsilon_0 R}\,.
\end{equation}
Figure \ref{fig:nonlinscreen} shows that nonlinear screening and linear screening predict different decay rates for the potential $\phi$ at short distances but have the same long-range decay rate. Since the screening near the origin is larger in the nonlinear case, the long-distance potential $\phi$ does not approach the weak field limit but is offset. As \reff{fig:nonlinscreen} demonstrates this is especially true in the BBN temperature range. All solutions have an exponential slope $-m_D$ indicating linear or weak field behavior at large distances.

The important prediction of the nonlinear screening in \reff{fig:nonlinscreen} is a rapid drop of the potential at short distances, thus a strong screening effect near the classical turning point of thermal reacting ions, where quantum tunneling for thermonuclear reactions will be important. This effect increases nonlinearly with $Z$, see \req{eq:nonlinscreen}. Note that the use of relativistic Fermi statistics may be required since the tunneling regime involves potentials that are comparable to and larger than the electron mass. We expect that nonlinear screening effects could lead to drastic changes in the Coulomb penetration factor for $Z>1$ and, in turn, light element abundances. 

To describe nonlinear screening beyond the presented toy model, in the context of light element screening during BBN, we need to solve \req{eq:Poisson-Boltz} with a realistic spherical charge distribution. That way the difference in asymptotic behavior will be captured for the solution at large distances. It turns out that even the simple first look at the 3D-static asymptotic Debye-type solution requires formidable numerical effort, let alone merging the short-range strong field limit with damped-dynamic linear response theory as developed here. Therefore the strong field screening will be addressed separately, opening the path to the computation of screened-BBN reaction rates.

%%%%%%%%%%%%%%%%%%%%%%%%%%%%%%%%%%%%%%%%%%%%
\section{Conclusion and discussion}\label{sec:Discussion}

This study sheds light on the properties and dynamics of the electron-positron plasma in the early Universe. In particular, we examined the damping rate and the electron-positron to baryon density ratio and their potential implications for Big Bang Nucleosynthesis (BBN) through screening within linear response theory. 

Our results confirm that BBN occurred in a dense $e^-e^+\gamma$ QED plasma interspersed with the heavy charged dust of protons and light nuclei. The polarization cloud surrounding an ionized nucleus reduces the Coulomb barrier needed to be overcome for a thermonuclear reaction to proceed. This increase in the penetration probability enhances the thermonuclear reaction rates. 

We calculated the screening effect on the internuclear Coulomb potential in the electron-positron plasma, accounting for the thermal motion of the nuclei and collisional damping. Our use of the current conserving collisional linear response framework for relativistic plasmas detailed in \cite{Formanek:2021blc} allowed us to present an approximate analytic formula in \req{eq:pos_point_DDS} to describe screening effects which were previously found only numerically. Our analytic formula can be readily used to estimate the effect of screening on thermonuclear reactions. Our results are illustrated in Fig.~\ref{fig:numericalComp}.

Common theoretical BBN models assume that the Universe was essentially void of anything except for reacting light nucleons and electrons needed to keep the local baryon density charge-neutral, a situation similar to the laboratory environment in which empirical nuclear reaction rates are obtained. Newer models, along with ours, address the BBN in dense electron-positron plasma using linear response approach \cite{Wang:2010px, Hwang:2021kno}. Since our results for damped-dynamic screening are very similar to dynamic screening found in~\cite{Hwang:2021kno} we expect that our work will not change existing predictions~\cite{Hwang:2021kno}, who obtained numerically the modification of Coulomb potential in linear response approximation but without damping. Here, we were able to obtain these results analytically, which helped us realize the limitations of the linear response approximation at short distances relevant for thermonuclear reactions. 

The prior and present study of BBN within the linear response approach has not fully recognized these limitations before: We believe that to model the BBN reaction network processes in dense primordial plasma, we need to obtain the strong screening effect when the potential strength is close to the thermal energy $|\phi|/T \approx 1$. This requires developing novel theoretical tools beyond the scope of this work. We showed, see \reff{fig:nonlinscreen}, that for the simple one-dimensional toy model, strong field screening could alter particle nuclear reaction rates, with importance increasing with $Z$. Since this will impact the BBN element production we will return to strong screening at short internuclear distances in follow-up work. Consideration of thermonuclear reaction rate enhancement estimation will follow the study of the short distance screening potential in the strong field limit. 

We currently calculate the damping strength to be much larger than the screening mass. This implies another necessary improvement in evaluating the strength of damping, which itself depends on plasma properties through the screening of scattering potential acting between charged plasma components. In forthcoming work, we will propose an approach where damping and plasma properties are determined in a self-consistent manner. We expect a self-consistent calculation to yield a smaller value as the damping effect reduces the magnitude of in plasma scattering cross-sections. The behavior of the plasma can alter significantly if the damping parameter becomes small enough that plasma oscillations become undamped.

Our study mirrors the theoretical findings of planetary and space dusty plasma theory~\cite{Montgomery:1970jpp, Stenflo:1973, Shukla:2002ppcf, Lampe:2000pop}. The large distance behavior of the damped-dynamic BBN screened potential in \req{eq:pos_point_DDS} matches that of slowly moving dust particles in plasma \cite{Stenflo:1973}. Dusty plasma theory studies a number of effects that are not currently included in BBN plasma studies, including dust charging, dust acoustic waves, dust instabilities, and structure formation (dust crystals)~\cite{Shukla:2002ppcf}. We expect that these results can be ported to the nuclear light element dust dynamics in the primordial $e^-e^+\gamma$ QED plasma. This interdisciplinary connection, which has not been recognized previously, could have substantial implications for our understanding of the evolution of matter in the early Universe, implying in particular primordial matter clumping.
 
It is compelling to consider that this polarization binding effect could lead to the early seeding of matter structure in the Universe. This happens due to the existence of an attractive portion of the internuclear (charged dust) potential which favors a specific average separation between charged dust particles. Therefore as the density of the dust in the Universe decreases due to Universe expansion, density inhomogeneity formation can be expected: The energetically favorable configuration consists of domains with `optimal' density of nuclear dust and other domains with reduced density or even entirely devoid of dust. Such inhomogeneities could influence the production of heavier isotopes and elements in the late stage of the BBN.

Dust crystals are known to form in polarized plasma~\cite{Shukla1996,Thomas:1994prl}. In our case nuclear `dust' clustering driven by in plasma polarization forces may provide a natural explanation for the recently discovered primordial galaxy accompanied by a large central black hole~\cite{Haro:2023JWST} existent when the Universe was barely 500 million years old. This is in our opinion suggestive of primordial matter concentration rather than evidence for other explanations, such as dark matter stars~\cite{Ilie:2023DM} or a much longer Universe lifespan~\cite{Gupta:2023mn} which could impact other predictions of the $\Lambda$CDM model~\cite{Sabti:2023xwo}. We believe that developing novel BBN plasma tools necessary for a screened nuclear reaction network, and a study of associated matter clumping in plasma driven by polarization forces could contribute to an improved understanding of JWST-HST spectacular observations.
 
Many of our results can be ported to other types of plasma than considered here. An immediate interdisciplinary application for this work could be the environment of stellar and `laboratory' plasma fusion with significant effects expected for the proton-Boron environment of intense current interest~\cite{Labaune:2013dla,Margarone:2022mdpi}. Another vibrant field of current research in acute need of a better understanding is the production of light nuclei in relativistic heavy ion collisions. These are emerging from high-temperature hadronic fireballs comprising predominantly of light pions and a few heavy `dust' nucleons at an ambient temperature vastly exceeding the nuclear binding. Their relatively abundant formation lacks a convincing explanation~\cite{Buyukcizmeci:2023azb}. In the same experimental relativistic heavy ion collision environment, the new strongly interacting quark-gluon plasma is formed initially. The gluon and light quark $u,d,s$ pair-plasma is seeded with heavy $c,\bar c, b, \bar b$ `charged dust'~\cite{Zhao:2020jqu}. This situation is also reminiscent of the environment explored here. However, additional complexity is introduced by the non-Abelian nature of strong interaction. 

In summary, our study has provided advances in understanding electron-positron plasma in the early Universe and its potential influence on the BBN. We discussed relation to other domains of physics where similar methods were used and presented several interdisciplinary directions for future research. Additionally, we have clarified steps going beyond the linear response method, which are necessary to understand the implications of our findings for BBN thermonuclear reaction rates and the evolution of the early Universe. 

\section*{Acknowledgements}
This research did not receive any specific grant from funding agencies in the public, commercial, or not-for-profit sectors. However, we thank Tam\'as Bir\'o for his hospitality at the PP2023 Margaret Island Symposium where this work was presented. An allocation of computer time was used from the High Performance Computing (HPC) resources supported by the University of Arizona TRIF, UITS, and Research, Innovation, and Impact (RII) and maintained by the UArizona Research Technologies department.
%%%%%%%%%%%%%%%%%%%%%%%%%%%%%%%%%%%%%%%%%%%%%%%%%%%%%%%%%%%%%%%%%%
%===================================================================
%===================================APPENDICES======================
\appendix
%===================================================================

\section{External charge distribution}\label{sec:freechg}

Here we define the free charge density used to describe light nuclei in BBN. We wish to model a nucleus moving at constant velocity $\boldsymbol{\beta}_N$. For simplicity, we model the charge distribution as a Gaussian in all directions
\begin{equation}
\rho_{\text{ext} }(t,\boldsymbol{x}) = \frac{Ze\gamma}{\pi^{3/2}R^3}e^{-\frac{1}{R^2}(\boldsymbol{x}-\boldsymbol{\beta}_N t )^2}\,,
\end{equation}
Since we work in the non-relativistic limit we assume that the Lorentz factor $\gamma\approx1$. The normalization is chosen in such a way that 
\begin{equation}
\int \rho_{\text{ext}}(t,\boldsymbol{x}) d^3\boldsymbol{x} = Ze\,
\end{equation}
is the total charge of the nucleus. The Gaussian radius parameter $R$ is related to the mean squared radius of the nucleus, $\langle r^2\rangle$ at rest, by
\begin{equation}\label{eq:radius}
\langle r^2 \rangle = \frac{1}{Ze}\int r^2 \rho_{\text{ext}}(\boldsymbol{x}) d^3\boldsymbol{x} = \frac{3}{2}R^2\,.
\end{equation}
Experimental measurements of the mean square radius can be found in Ref.~\cite{DeVries:1987atn}. The Fourier transformed charge distribution is
\begin{equation}\label{eq:extchgfreq}
\wt{\rho}_{\text{ext}}(\omega,\boldsymbol{k}) = 2\pi Ze\, e^{-\boldsymbol{k}^2 \frac{R^2}{4}} \delta(\omega - \boldsymbol{k}\cdot \boldsymbol{\beta}_N)\,,
\end{equation} 
where $\delta$ is the Dirac delta function in one dimension.

\section{Static screening for a Gaussian charge distribution}\label{sec:static}
Starting from \req{eq:potentk} we can take the limit $\beta_N\rightarrow0$ to find the static screening potential for a Gaussian charge distribution
\begin{equation}
 \phi_{\text{stat}}(t,\boldsymbol{x}) = Ze\int \frac{d^3\boldsymbol{k}}{(2\pi)^3} e^{ i\boldsymbol{k}\cdot \boldsymbol{x}} \frac{ e^{-\boldsymbol{k}^2\frac{R^2}{4}}}{\boldsymbol{k}^2+m_D^2}\,.
\end{equation}
First, we perform the angular integration to find
\begin{equation}
 \phi_{\text{stat}}(t,\boldsymbol{x}) = Ze\int_0^{\infty} \frac{k^2 d{k}}{(2\pi)^2} \frac{2\sin{({k}|\boldsymbol{x}|)}}{{k}|\boldsymbol{x}|} \frac{ e^{-{k}^2\frac{R^2}{4}}}{{k}^2+m_D^2}\,.
\end{equation}
This integral can be evaluated using a Schwinger parameterization
\begin{equation}\label{eq:schwin}
 \phi_{\text{stat}}(t,\boldsymbol{x}) = Ze\int_0^{\infty} ds \int_0^{\infty} \frac{k^2 d{k}}{(2\pi)^2} \frac{2\sin{({k}|\boldsymbol{x}|)}}{{k}|\boldsymbol{x}|} e^{-{k}^2\frac{R^2}{4}-s({{k}^2+m_D^2})}\,,
\end{equation}
where we first perform the integration over $k$. The $k$ integration is calculated by writing the $\sin$ in its exponential form and splitting it into two terms with opposite signs in the exponents. These two terms can be recombined into one Fourier transform from $-\infty$ to $+\infty$. Finally, the integration over $s$ results in the position space static screening potential
\begin{multline}\label{eq:Stat_Gauss}
 \phi_{\text{stat}}(t,\boldsymbol{x}) =\frac{Z e}{4 \pi \varepsilon_0}\frac{e^{m_D^2 R^2/4}}{r}\Bigg(\frac{ \text{Erf}\left(\frac{r}{R} - \frac{m_D R}{2}\right)}{2 }e^{-m_D r} \\+\frac{ \text{Erf}\left(\frac{r}{R} + \frac{m_D R}{2}\right)}{2 }e^{m_D r} - \sinh(m_D r)\Bigg)\,,
\end{multline}
where we have reintroduced the vacuum permittivity $\varepsilon_0$. Taking the limit of this expression for $R\rightarrow0$ yields the usual static screening result \cite{Debye:1923}
\begin{equation}
 \lim_{R\rightarrow 0} \phi_{\text{stat}}(t,\boldsymbol{x}) = \frac{Z e}{4 \pi \varepsilon_0}\frac{e^{-m_D r}}{r}\,.
\end{equation}

%\newpage

%%%%%%%%%%%%%%%%%%%%%%%%%%%%%%%%%%%%%%%%%%%%%%%%%%%%%%%%%%%%%%%%%%%%%%

%%%%%%%%%%%%%%%%%%%%%%%%%%%%%%%%%%%%%%%%%%%%%%%%%%%%%%%%%%%%%%%%%%%
\end{document}